\documentclass[12pt]{article}
\usepackage{amsfonts,amsmath,graphicx,setspace}
\usepackage{color,hyperref,verbatim,natbib,rotating}

\definecolor{Red}{rgb}{0.5,0,0}
\definecolor{Blue}{rgb}{0,0,0.5}
\hypersetup{%
    colorlinks = {true},
    linktocpage = {true},
    plainpages = {false},
    linkcolor = {Blue},
    citecolor = {Blue},
    urlcolor = {Red},
    pdfstartview = {XYZ null null 1.25},
    pdfpagemode = {UseOutlines},
    pdfview = {XYZ null null null}
}
\newcommand{\email}[1]{\href{mailto:#1}{\normalfont\texttt{#1}}}
\newcommand{\R}{\textsf{R}}

\newcommand{\bfalpha}{\mbox{{\boldmath $\alpha$}}}
\newcommand{\bfbeta}{\mbox{{\boldmath $\beta$}}}
\newcommand{\bfgamma}{\mbox{{\boldmath $\gamma$}}}
\newcommand{\bftheta}{\mbox{{\boldmath $\theta$}}}

\newcommand{\eps}{\varepsilon}
\newcommand{\bw}{\mbox{{\boldmath $w$}}}
\newcommand{\bv}{\mbox{{\boldmath $v$}}}
\newcommand{\by}{\mbox{{\boldmath $y$}}}
\newcommand{\bx}{\mbox{{\boldmath $x$}}}

\newcommand{\bz}{\mbox{{\boldmath $z$}}}

\newcommand{\bb}{\mbox{{\boldmath $b$}}}

\newcommand{\bD}{\mbox{{\boldmath $D$}}}

\bibpunct{(}{)}{;}{a}{}{,}

\begin{document}

{\vspace*{1cm}}

\begin{center}
\Large \bf Combining Dynamic Predictions from Joint Models for Longitudinal and Time-to-Event Data using Bayesian Model Averaging
\end{center}

\vspace{0.5cm}
%\begin{comment}
\begin{center}
{\large Dimitris Rizopoulos$^{1,*}$, Laura A. Hatfield$^{2}$, Bradley P. Carlin$^{3}$ and Johanna J.M. Takkenberg$^{4}$}\footnote{$^*$Correspondance at: Department of Biostatistics, Erasmus University Medical Center, PO Box 2040, 3000 CA Rotterdam, the Netherlands. E-mail address: \email{d.rizopoulos@erasmusmc.nl}.}\\
$^{1}$ Department of Biostatistics, Erasmus University Medical Center\\
$^{2}$ Department of Health Care Policy, Harvard Medical School\\
$^{3}$ Division of Biostatistics, School of Public Health, University of Minnesota\\
$^{4}$ Department of Cardiothoracic Surgery, Erasmus University Medical Center\\
\end{center}
%\end{comment}
\vspace{0.6cm}

%=====================================================

\begin{spacing}{1}
\noindent {\bf Abstract}\\
The joint modeling of longitudinal and time-to-event data is an active area of statistics research that has received a lot of attention in the recent years. More recently, a new and attractive application of this type of models has been to obtain individualized predictions of survival probabilities and/or of future longitudinal responses. The advantageous feature of these predictions is that they are dynamically updated as extra longitudinal responses are collected for the subjects of interest, providing real time risk assessment using all recorded information. The aim of this paper is two-fold. First, to highlight the importance of modeling the association structure between the longitudinal and event time responses that can greatly influence the derived predictions, and second, to illustrate how we can improve the accuracy of the derived predictions by suitably combining joint models with different association structures. The second goal is achieved using Bayesian model averaging, which, in this setting, has the very intriguing feature that the model weights are not fixed but they are rather subject- and time-dependent, implying that at different follow-up times predictions for the same subject may be based on different models.\\\\
\noindent {\it Keywords:} Prognostic Modeling; Risk Prediction; Random Effects; Time-Dependent Covariates.
\end{spacing}

%=====================================================

\section{Introduction} \label{Sec:Intro}
In recent years it has been recognized that personalized medicine forms the future of medical care. This has increased interest in development of prognostic models for many different types of diseases. Examples are numerous and include, prognostic models for various types of cancer, such as breast and prostate cancer; risk scores for cardiovascular diseases, such as the Framingham score; and prognostic models applied in AIDS research to assess the risk of HIV infected patients. However, even though there is a wealth of patient data available, the majority of prognostic models in the literature provide risk predictions using only a small portion of the recorded information. This is especially true for patient outcomes that are repeatedly measured in time where typically only the last measurement is utilized. A clear advantage of such simple models is that they can be applied in everyday clinical practice. However, an important limitation is that valuable information is discarded, which if appropriately used, could offer a better insight into the dynamics of the disease's progression. In particular, an inherent characteristic of many medical conditions, such as those described above, is their dynamic nature. That is, the rate of progression is not only different from patient to patient but also dynamically changes in time for the same patient. Hence, it is medically relevant to investigate whether repeated measurements of a biomarker can provide a better understanding of disease progression and a better prediction of the risk for the event of interest than a single biomarker measurement (e.g., at baseline or the last available). It is evident that markers with this capability would become a valuable tool in everyday medical practice, because they would provide physicians with a better understanding of disease progression for a particular patient, and allow them to make more informed decisions. This is also the aim of our motivating case study on patients who underwent aortic valve allograft implantation, described in detail in Section~\ref{Sec:AoValvInf}. In particular, even though human tissue valves have some advantages, they are also more susceptible to tissue degeneration and require re-intervention. Hence, cardiologists wish to use an accurate prognostic tool that will inform them about the future prospect of a patient with a human tissue valve in order to adjust medical care and postpone re-operation and/or death.

Motivated by the Aortic Valve case study, this paper aims to provide a flexible modeling framework for producing risk predictions that utilize all available patient information. We will explicitly model the longitudinal history of each subject by basing our developments on the framework of joint models for longitudinal and time-to-event data \citep{faucett.thomas:96, wulfsohn.tsiatis:97, henderson.et.al:00, tsiatis.davidian:04, guo.carlin:04, rizopoulos:12}. In particular, an attractive use of joint models is to derive dynamic predictions for either the survival or longitudinal outcomes \citep{yu.et.al:08, proust-lima.taylor:09, rizopoulos:11}. The advantageous features of these predictions are that they are individualized, are updated as extra longitudinal information is recorded for each subject, and are based on all past values of the longitudinal outcome. Our novel contributions are two-fold. First, the subjects under consideration often exhibit complex longitudinal trajectories with nonlinearities and plateaus. In these settings, it is relevant to consider which characteristics of a subject's trajectory best predict the event of interest. To this end, we will investigate how predictions are affected by assuming different types of association structure between the longitudinal and event time processes. We go beyond the standard formulation of joint models \citep{henderson.et.al:00}, and we postulate functional forms that allow the rate of increase/decrease of the longitudinal outcome or a suitable summary of the whole longitudinal trajectory to determine the risk for an event. The consideration of competing association structures to describe the link between the two processes raises the issue of model uncertainty, which is typically ignored in prognostic modeling. Motivated by this, our second contribution is to derive predictions based not on a single model but on a collection of models simultaneously, combining them using Bayesian model averaging. As we will show later this approach bases predictions are based on the available data of a subject and the model weights are both individual- and time-dependent.

The rest of the paper is organized as follows. Section~\ref{Sec:AoValvInf} gives a background on aortic allograft implantation and describes the Aortic Valve dataset that motivates this research. Section~\ref{Sec:JM} briefly introduces the joint modeling framework, presents estimation under a Bayesian approach and shows how dynamic individualized predictions can be derived under a joint model. Section~\ref{Sec:Parms} introduces several formulations of the association structure between the longitudinal and survival processes. Section~\ref{Sec:BMA} presents the Bayesian model averaging methodology to combine predictions, and Section~\ref{Sec:Appl} illustrates the use of this technique in the Aortic Valve dataset. Section~\ref{Sec:Simul} refers to the results of a simulation study, and Section~\ref{Sec:Disc} concludes the paper.

%=====================================================

\section{Background on the Aortic Valve Dataset} \label{Sec:AoValvInf}
The motivation for this research comes from a study, conducted by the Department of Cardio-Thoracic Surgery of the Erasmus Medical Center in the Netherlands that includes 286 patients who received a human tissue valve in the aortic position in the hospital from 1987 until 2008 \citep{bekkers.et.al:11}. Aortic allograft implantation has been widely used for a variety of aortic valve or aortic root diseases. Initial reports on the use of either fresh or cryopreserved allografts date from the early years of heart valve surgery. Major advantages ascribed to allografts are the excellent hemodynamic characteristics as a valve substitute; the low rate of thrombo-embolic complications, and, therefore, absence of the need for anticoagulant treatment; and the resistance to endocarditis. A major disadvantage of using human tissue valves, however is the susceptibility to (tissue) degeneration and need for re-interventions. The durability of a cryopreserved aortic allograft is age-dependent, leading to a high lifetime risk of re-operation, especially for young patients. Re-operations on the aortic root are complex, with substantial operative risks, and mortality rates in the range 4--12\%.

\begin{figure}[!h]
\includegraphics[width = \textwidth]{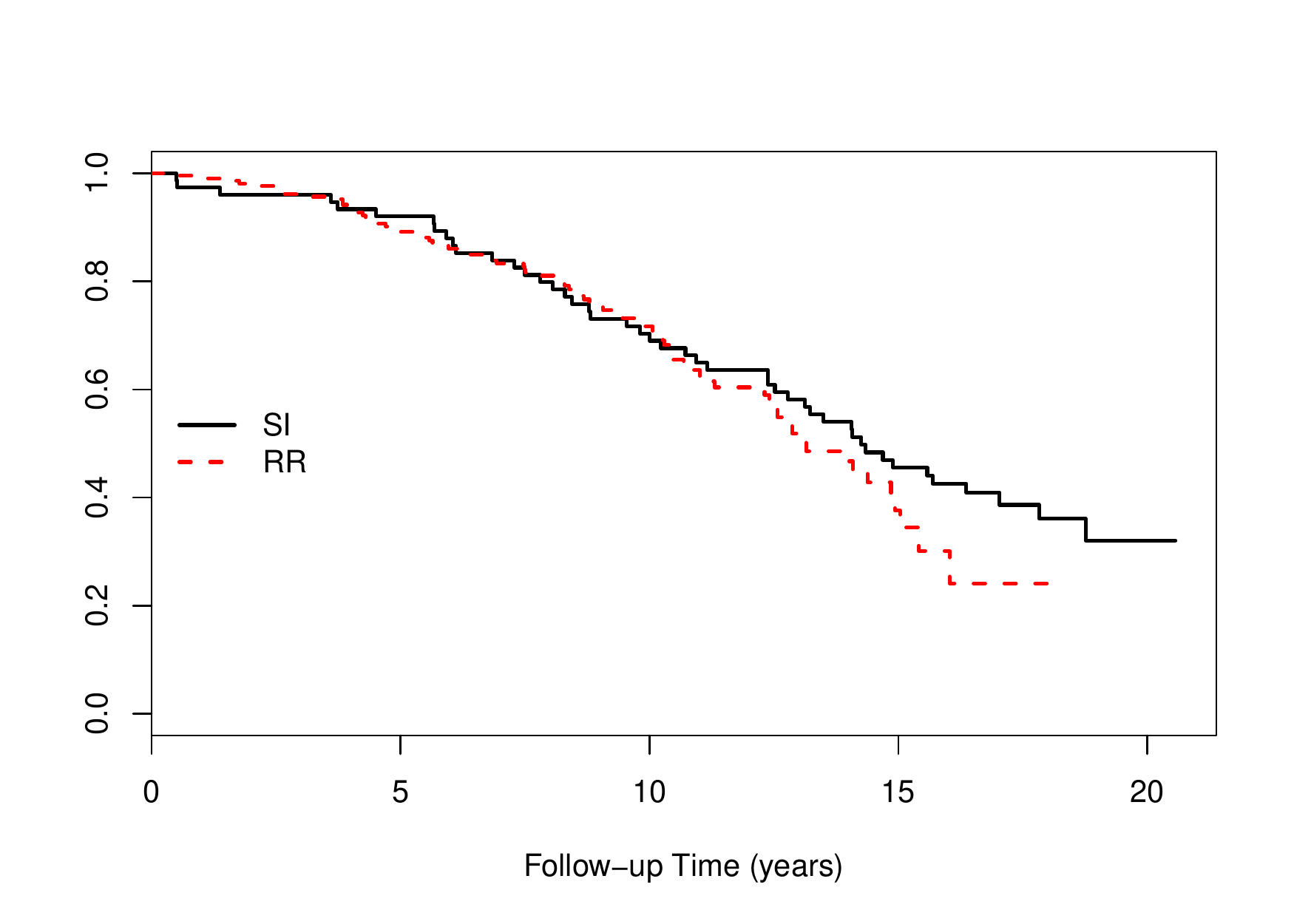}
\caption{Kaplan-Meier estimates of the survival functions for re-operation-free survival for the sub-coronary implantation (SI) and root replacement (RR) groups.} \label{Fig:KM}
\end{figure}

\begin{figure}[!h]
\includegraphics[width = \textwidth]{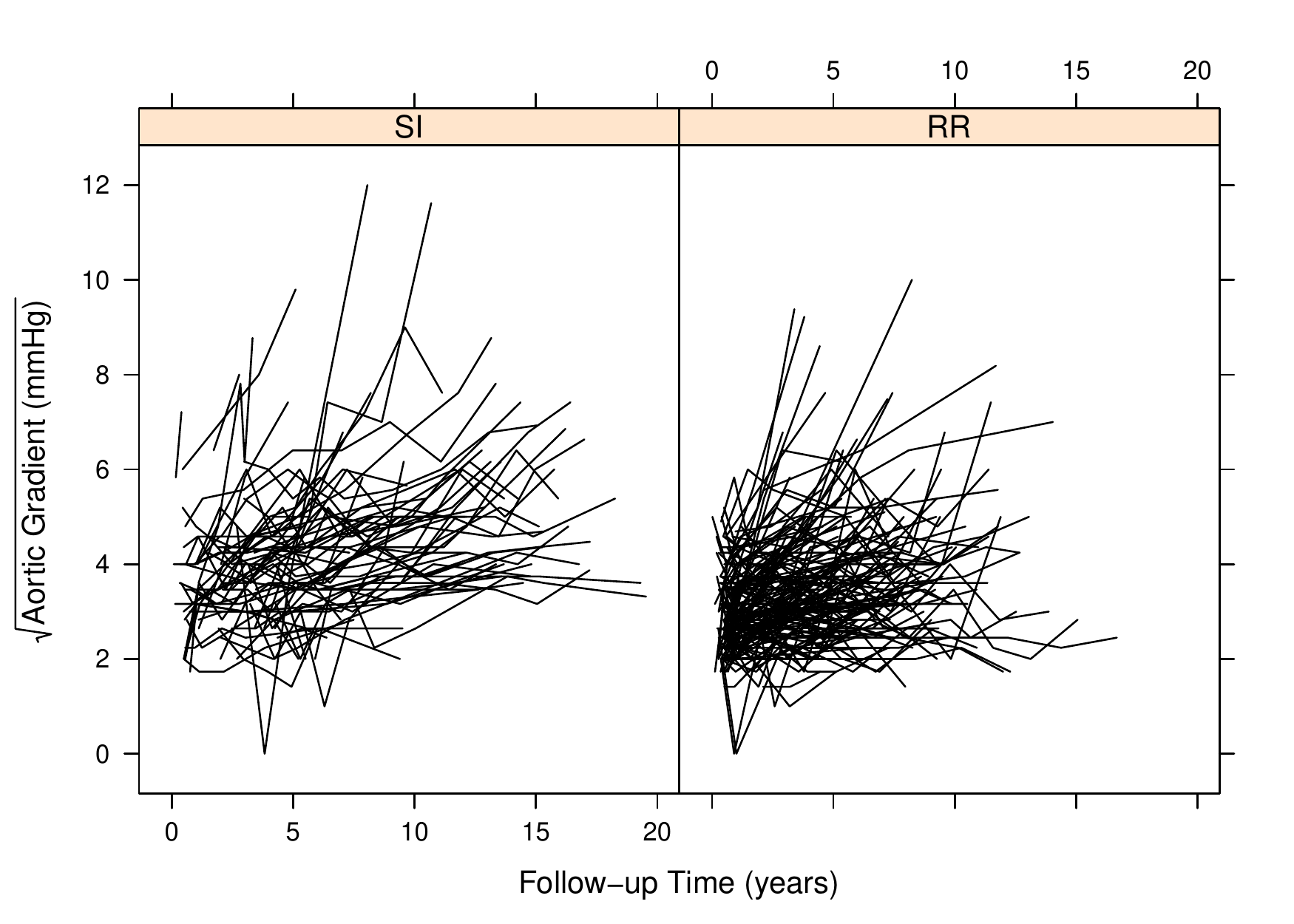}
\caption{Subject-specific profiles for the square root aortic gradient separately for the sub-coronary implantation (SI) and root replacement (RR) groups.} \label{Fig:SubjProfiles}
\end{figure}

In our study, a total of 77 (26.9\%) patients received a sub-coronary implantation (SI) and the remaining 209 patients a root replacement (RR). These patients were followed prospectively over time with annual telephone interviews and biennial standardized echocardiographic assessment of valve function until July 8, 2010. Echo examinations were scheduled at 6 months and 1 year postoperatively, and biennially thereafter, and at each examination, echocardiographic measurements of aortic gradient (mmHg) were taken. By the end of follow-up, 1241 aortic gradient measurements were recorded with an average of 5 measurements per patient (s.d. 2.3 measurements), 59 (20.6\%) patients had died, and 73 (25.5\%) patients required a re-operation on the allograft. Following the discussion in Section~\ref{Sec:Intro}, our aim here is to use the existing data to construct a prognostic tool that will provide accurate risk predictions for future patients from the same population, utilizing their baseline information, namely age, gender and the type of operation they underwent, and their recorded aortic gradient levels. The composite event re-operation or death was observed for 125 (43.7\%) patients, and the corresponding Kaplan-Meier estimator for the two intervention groups is shown in Figure~\ref{Fig:KM}.
We can observe minimal differences in the re-operation-free survival rates between sub-coronary implantation and root replacement, with only a slight advantage of sub-coronary implantation towards the end of the follow-up. For the longitudinal outcome, Figure~\ref{Fig:SubjProfiles} depicts the subject-specific longitudinal profiles for the two intervention groups.
Because aortic gradient exhibits right skewness, we use the square root transform of aortic gradient. We observe considerable variability in the shapes of these trajectories, but there are no systematic differences apparent between the two groups.

%=====================================================

\section{Joint Model Specification and Predictions} \label{Sec:JM}
\subsection{Submodels} \label{Sec:JM-submodels}
Let $T_i^*$ denote the true event time for the $i$-th subject ($i = 1, \ldots, n$), $C_i$ the censoring time, and $T_i = \min(T_i^*, C_i)$ the corresponding observed event time. In addition, we let $\delta_i = I(T_i^* \leq C_i)$ denote the event indicator, with $I(\cdot)$ being the indicator function that takes the value 1 when $T_i^* \leq C_i$, and 0 otherwise. For the longitudinal process, we let $\by_i$ denote the $n_i \times 1$ longitudinal response vector for the $i$-th subject, with element $y_{il}$ denoting the value of the longitudinal outcome taken at time point $t_{il}$, $l = 1, \ldots, n_i$. Our aim is to postulate a suitable joint model that associates these two processes.

Focusing on normally distributed longitudinal outcomes, we use a linear mixed-effects model to describe the subject-specific longitudinal trajectories. Namely, we have
\begin{eqnarray}
\begin{array}{rcl}
y_i(t) & = & m_i(t) + \eps_i(t) = \bx_i^\top(t) \bfbeta + \bz_i^\top(t) \bb_i + \eps_i(t),\\
\bb_i & \sim & \mathcal N (\mathbf{0}, \bD), \quad \eps_i(t) \sim \mathcal N (0, \sigma^2),\\
\end{array} \label{Eq:LinearMixed}
\end{eqnarray}
where $y_i(t)$ denotes the value of the longitudinal outcome at any particular time point $t$, $\bx_i(t)$ and $\bz_i(t)$ denote the time-dependent design vectors for the fixed-effects $\bfbeta$ and for the random effects $\bb_i$, respectively, and $\eps_i(t)$ the corresponding error terms that are assumed independent of the random effects, and $\mbox{cov} \{\eps_i(t), \eps_i(t')\} = 0$ for $t' \neq t$. For the survival process, we assume that the risk for an event depends on the true and unobserved value of the marker at time $t$, denoted by $m_i(t)$ in \eqref{Eq:LinearMixed}. More specifically, we have
\begin{eqnarray}
\nonumber h_i (t \mid \mathcal M_i(t), \bw_i) & = & \lim_{s \rightarrow 0} \Pr \{ t \leq T_i^* < t + s \mid T_i^* \geq t, \mathcal M_i(t), \bw_i \} \big / s\\
& = & h_0(t) \exp \bigl \{ \bfgamma^\top
\bw_i + \alpha m_i(t) \bigr \}, \quad t > 0, \label{Eq:Surv-RR}
\end{eqnarray}
where $\mathcal M_i(t) = \{ m_i(s), 0 \leq s < t \}$ denotes the history of the true unobserved longitudinal process up to $t$, $h_0(\cdot)$ denotes the baseline hazard function, and $\bw_i$ is a vector of baseline covariates with corresponding regression coefficients $\bfgamma$. Parameter $\alpha$ quantifies the association between the true value of the marker at $t$ and the hazard for an event at the same time point. To complete the specification of the survival process we need to make appropriate assumptions for the baseline hazard function $h_0(\cdot)$. Typically in survival analysis this is left unspecified. However, since our aim is to make subject-specific predictions of survival probabilities, we use a smoother assumption for this function. We achieve this, while still allowing for flexibility in the specification of $h_0(\cdot)$ by using a B-splines approach. In particular, the log baseline hazard function is expressed as
\begin{equation}
\log h_0(t) = \gamma_{h_0,0} + \sum \limits_{q = 1}^Q \gamma_{h_0,q} B_q(t, \bv), \label{Eq:BaseHaz}
\end{equation}
where $B_q(t, \bv)$ denotes the $q$-th basis function of a B-spline with knots $v_1, \ldots, v_Q$ and $\bfgamma_{h_0}$ the vector of spline coefficients. Increasing the number of knots $Q$ increases the flexibility in approximating $\log h_0(\cdot)$; however, we should balance bias and variance and avoid overfitting. A standard rule of thumb is to keep the total number of parameters, including the parameters in the linear predictor in \eqref{Eq:Surv-RR} and in the model for $h_0(\cdot)$, between 1/10 and 1/20 of the total number of events in the sample \citep[Section~4.4]{harrell:01}. After the number of knots has been decided, their location is based on percentiles of the observed event times $T_i$ or of the true event times $\{T_i : T_i^* \leq C_i, i = 1, \ldots, n\}$ to allow for more flexibility in the region of greatest density.

%%%%%%%%%%%%%%%

\subsection{Estimation} \label{Sec:JM-estimation}
For the estimation of the joint model's parameters we follow the Bayesian paradigm and derive posterior inferences using a Markov chain Monte Carlo algorithm (MCMC). The likelihood of the models is derived under the assumption that the vector of time-independent random effects $\bb_i$ accounts for all interdependencies between the observed outcomes. That is, given the random effects, the longitudinal and event time process are assumed independent, and in addition, the longitudinal responses of each subject are assumed independent. Formally we have,
\begin{eqnarray}
p(\by_i, T_i, \delta_i \mid \bb_i, \bftheta) & = & p(\by_i \mid \bb_i, \bftheta) \; p(T_i, \delta_i \mid \bb_i, \bftheta), \label{Eq:CondInd-I}\\
p(\by_i \mid \bb_i, \bftheta) & = & \prod_l p ( y_{il} \mid \bb_i, \bftheta ), \label{Eq:CondInd-II}
\end{eqnarray}
where $\bftheta^\top = (\bftheta_t^\top, \bftheta_y^\top, \bftheta_b^\top)$ denotes the full parameter vector, with $\bftheta_t$ denoting the parameters for the event time outcome, $\bftheta_y$ the parameters for the longitudinal outcomes, and $\bftheta_b$ the unique parameters of the random-effects covariance matrix, and $p(\cdot)$ denotes an appropriate probability density function. In addition, we assume that given the observed history of longitudinal responses up to time $s$, the censoring mechanism and the visiting process are independent of the true event times and future longitudinal measurements at $t > s$. Under these assumptions, the likelihood contribution for the $i$-th subject conditional on the parameters and random effects takes the form:
\begin{eqnarray}
\lefteqn{p(\by_i, T_i, \delta_i \mid \bftheta, \bb_i) = \prod \limits_{l=1}^{n_i} p ( y_{il} \mid \bb_i; \bftheta_y ) \, p(T_i, \delta_i \mid \bb_i; \bftheta_t, \bfbeta) \, p(\bb_i; \bftheta_b) \label{Eq:Log-Lik}}\\
\nonumber & \propto & \biggl [
(\sigma^2)^{-n_i/2} \exp \Bigl \{ - \sum_l \bigl (y_{il} - \bx_{il}^\top \bfbeta - \bz_{il}^\top \bb_i \bigr )^2 / 2 \sigma^2  \Bigr \}\\
\nonumber & \times &
\Bigl [ \exp \Bigl \{ \sum_q \gamma_{h_0,q} B_q(T_i, \bv) + \bfgamma^\top \bw_i + \alpha m_i(T_i) \Bigr \} \Bigr ]^{\delta_i}\\
\nonumber & \times & \exp \Bigl [ - \exp(\bfgamma^\top \bw_i) \int_0^{T_i} \exp \Bigl \{\sum_q \gamma_{h_0,q} B_q(s, \bv) + \alpha m_i(s) \Bigr \} \, ds \Bigr ]\\
\nonumber & \times & \mbox{det}(\bD)^{-1/2} \exp \bigl (- \bb_i^\top \bD^{-1} \bb_i / 2 \bigr) \biggr ],
\end{eqnarray}
where the intercept term $\gamma_{h_0,0}$ from the definition of the baseline risk function \eqref{Eq:BaseHaz} has been incorporated into the design vector $\bw_i$.  The integral in the definition of the survival function
\begin{equation}
S_i(t \mid \mathcal M_i(t), \bw_i) = \exp \Bigl \{- \int_0^t h_0(s) \exp \bigl \{ \bfgamma^\top \bw_i + \alpha m_i(s) \bigr \} ds \Bigr \}, \label{Eq:SurvivalFun}
\end{equation}
does not have a closed-form solution, and thus a numerical method must be employed for its evaluation. Here we use a 15-point Gauss-Kronrod quadrature rule. For the parameters $\bftheta$ we take standard prior distributions. In particular, for the vector of fixed effects of the longitudinal submodel $\bfbeta$, for the regression parameters of the survival model $\bfgamma$, for the vector of spline coefficients for the baseline hazard $\bfgamma_{h_0}$, and for the association parameter $\alpha$ we use independent univariate diffuse normal priors. For the variance of the error terms $\sigma^2$ we take an inverse-Gamma prior, while for covariance matrices we assume an inverse Wishart prior.

%%%%%%%%%%%%%%%

\subsection{Dynamic Individualized Predictions} \label{Sec:DynPred}
Under the Bayesian specification of the joint model, presented in Section~\ref{Sec:JM}, we can derive subject-specific predictions for either the survival or longitudinal outcomes \citep{yu.et.al:08, rizopoulos:11, rizopoulos:12}. To put it more formally, based on a joint model fitted in a sample $\mathcal D_n = \{T_i, \delta_i, \by_i; i = 1, \ldots, n\}$ from the target population, we are interested in deriving predictions for a new subject $j$ from the same population that has provided a set of longitudinal measurements $\mathcal Y_j(t) = \{ y_j(s); 0 \leq s \leq t \}$, and has a vector of baseline covariates $\bw_j$. The fact that biomarker measurements have been recorded up to $t$, implies survival of this subject up to this time point, meaning that it is more relevant to focus on the conditional subject-specific predictions, given survival up to $t$. In particular, for any time $u > t$ we are interested in the probability that this new subject $j$ will survive at least up to $u$, i.e.,
\[
\pi_j(u \mid t) = \Pr ( T_j^* \geq u \mid T_j^* > t, \mathcal Y_j(t), \bw_j, \mathcal D_n).
\]
Similarly, for the longitudinal outcome we are interested in the predicted longitudinal response at $u$, i.e.,
\[
\omega_j(u \mid t) = E \bigl \{ y_j(u) \mid T_j^* > t, \mathcal Y_j(t), \mathcal D_n \bigr \}.
\]
The time-dynamic nature of both $\pi_j(u \mid t)$ and $\omega_j(u \mid t)$ is evident because when new information is recorded for patient $j$ at time $t' > t$, we can update these predictions to obtain $\pi_j(u \mid t')$ and $\omega_j(u \mid t')$, and therefore proceed in a time dynamic manner.

Under the joint modeling framework of Section~\ref{Sec:JM}, estimation of either $\pi_j(u \mid t)$ or $\omega_j(u \mid t)$ is based on the corresponding posterior predictive distributions, namely
\[
\pi_j(u \mid t) = \int \Pr(T_j^* \geq u \mid T_j^* > t, \mathcal Y_j(t), \bftheta) \, p(\bftheta \mid \mathcal D_n) \, d\bftheta,
\]
for the survival outcome, and analogously
\[
\omega_j(u \mid t) = \int E \bigl \{ y_j(u) \mid T_j^* > t, \mathcal Y_j(t), \bftheta \bigr \} \, p(\bftheta \mid \mathcal D_n) \, d\bftheta,
\]
for the longitudinal one. The calculation of the first part of each integrand takes full advantage of the conditional independence assumptions \eqref{Eq:CondInd-I} and \eqref{Eq:CondInd-II}. In particular, we observe that the first factor of the integrand of $\pi_j(u \mid t)$ can be rewritten by noting that:
\begin{eqnarray*}
\nonumber \Pr(T_j^* \geq u \mid T_j^* > t, \mathcal Y_j(t), \bftheta) & = & \int \Pr(T_j^* \geq u \mid T_j^* > t, \bb_j, \bftheta) \, p(\bb_j \mid T_j^* > t, \mathcal Y_j(t), \bftheta) \, d\bb_j\\
& = & \int \frac{S_j \bigl \{ u \mid \mathcal M_j(u, \bb_j), \bftheta \bigr \}}{S_j \bigl \{ t \mid \mathcal M_j(t, \bb_j), \bftheta \bigr \} } \, p(\bb_j \mid T_j^* > t, \mathcal Y_j(t), \bftheta) \, d\bb_j,
\end{eqnarray*}
whereas for $\omega_j(u \mid t)$ we similarly have:
\begin{eqnarray*}
\nonumber E \bigl \{ y_j(u) \mid T_j^* > t, \mathcal Y_j(t), \bftheta \bigr \} & = & \int E \bigl \{ y_j(u) \mid \bb_j, \bftheta \} \,
p(\bb_j \mid T_j^* > t, \mathcal Y_j(t), \bftheta) \, d\bb_j\\
& = &  \bx_j^\top(u) \bfbeta + \bz_j^\top(u) \bar \bb_j^{(t)},
\end{eqnarray*}
with
\[
\bar \bb_j^{(t)} = \int \bb_j \, p(\bb_j \mid T_j^* > t, \mathcal Y_j(t), \bftheta) \, d\bb_j.
\]
Combining these equations with the MCMC sample from the posterior distribution of the parameters for the original data $\mathcal D_n$, we can devise a simple simulation scheme to obtain Monte Carlo estimates of $\pi_j(u \mid t)$ and $\omega_j(u \mid t)$. More details can be found in \citet{yu.et.al:08} and \citet{rizopoulos:11, rizopoulos:12}.

%=====================================================

\section{Functional Form} \label{Sec:Parms}
In ordinary proportional hazards models it has been long recognized that the functional form of time-varying covariates influences the derived inferences; see, for example, \citet{fisher.lin:99} and references therein. In the joint modeling framework however, where the longitudinal outcome plays the role of a time-dependent covariate for the survival process, this topic has received much less attention. The two main functional forms that have been primarily used so far in joint models include in the linear predictor of the relative risk model \eqref{Eq:Surv-RR} either the subject-specific means $m_i(t)$ from the longitudinal submodel or just the random effects $b_i$ \citep{henderson.et.al:00, rizopoulos.ghosh:11}. Nonetheless, in our setting, where our primary interest is in producing accurate predictions, we expect that the link between the longitudinal and event time processes to be important, and therefore it is relevant to investigate if and how the accuracy of predictions is influenced by the assumed functional form. This is motivated by the fact that there could be other characteristics of the patients' longitudinal profiles that are more strongly predictive for the risk of an event, such as the rate of increase/decrease of the biomarker's levels or a suitable summary of the whole longitudinal trajectory.

In this section, we present a few examples of alternative formulations for the association structure between the longitudinal outcome and the risk for an event. These different parameterizations can be seen as special cases of the following general formulation of the relative risk model:
\[
h_i(t) = h_0(t) \exp \bigr [ \bfgamma^\top \bw_i + \bfalpha^\top f \{t, \bb_i, \mathcal M_i (t)\} \bigr ],
\]
where $f(\cdot)$ is a time-dependent function that may depend on the random effects and on the true longitudinal history, as approximated by the mixed-effects model, and $\bfalpha$ is a vector of association parameters.

%%%%%%%%%%%%%%%

\subsection{Time-Dependent Slopes}
\label{Sec:ParmsTDslope}
\noindent The standard formulation \eqref{Eq:Surv-RR} postulates that the risk for an event at time $t$ is associated with parameter $\alpha$ measuring the strength of this association. Even though this is a very intuitively appealing parameterization with a clear interpretation for $\alpha$, it cannot distinguish between patients who, for instance, at a specific time point have equal true marker levels, but they may differ in the rate of change of the marker, with one patient having an increasing trajectory and the other a decreasing one. An extension of \eqref{Eq:Surv-RR} to capture such a setting has been considered by \citet{ye.et.al:08b} who posited a joint model in which the risk depends on both the current true value of the trajectory and the slope of the true trajectory at time $t$. More specifically, the relative risk survival submodel takes the form,
\begin{equation}
h_i(t) = h_0(t) \exp \bigl \{ \bfgamma^\top \bw_i + \alpha_1 m_i(t)  + \alpha_2 m_i'(t) \bigr \}, \label{Eq:Param-TDslopes}
\end{equation}
where $m_i'(t) = d\{\bx_i^\top (t) \bfbeta + \bz_i^\top (t) \bb_i\} / dt$. The interpretation of parameter $\alpha_1$ remains the same as in the standard parameterization \eqref{Eq:Surv-RR}. Parameter $\alpha_2$ measures the association between the value of the slope of the true longitudinal trajectory at time $t$ and the risk for an event at the same time point, provided that $m_i(t)$ remains constant.

%%%%%%%%%%%%%%%

\subsection{Cumulative Effects}
\label{Sec:ParmsCumEff}
\noindent A common characteristic of both \eqref{Eq:Surv-RR} and \eqref{Eq:Param-TDslopes} is that the risk for an event at any time $t$ is assumed to be associated with features of the longitudinal trajectory at the same time point. However, in ordinary time-dependent Cox models several authors have argued that this assumption is over-simplistic, and in many real settings we may benefit from allowing the risk to depend on a more elaborate function of the history of the time-varying covariate \citep{sylvestre.abrahamowicz:09}. Extending this concept in the context of joint models, we will account for the cumulative effect of the longitudinal outcome by including in the linear predictor of the relative risk submodel the integral of the longitudinal trajectory from baseline up to time $t$. More specifically, the survival submodel takes the form
\begin{equation}
h_i(t) = h_0(t) \exp \Bigl \{ \bfgamma^\top \bw_i + \alpha \int_0^t m_i(s) \, ds \Bigr \}, \label{Eq:Param-CumEff}
\end{equation}
where for any particular time point $t$, $\alpha$ measures the strength of the association between the risk for an event at time point $t$ and the area under the longitudinal trajectory up to the same time $t$, with the area under the longitudinal trajectory taken as a suitable summary of the whole marker history $\mathcal M_i(t) = \{ m_i(s), 0 \leq s < t \}$.

A feature of \eqref{Eq:Param-CumEff} is that it assigns the same weight on all past values of the longitudinal trajectory. This may not be reasonable when biomarker values closer to $t$ are considered more relevant. An extension that allows placing different weights at different time points is to multiply $m_i(t)$ with an appropriately chosen weight function $\varpi(\cdot)$ that places different weights at different time points, i.e.,
\begin{equation}
h_i(t) = h_0(t) \exp \Bigl \{ \bfgamma^\top \bw_i + \alpha \int_0^t \varpi(t - s) m_i(s) \, ds \Bigr \}. \label{Eq:Param-wCumEff}
\end{equation}
A possible family of functions with this property are probability density functions of known parametric distributions, such as the normal, the Student's-$t$ and the logistic. The scale parameter in these densities and the degrees of freedom parameter in the Student's-$t$ density can be used to tune the relative weights of more recent marker values compared to older ones.

%%%%%%%%%%%

\subsection{Shared Random Effects} \label{Sec:ParmsRE}
As mentioned earlier, one of the standard formulations of joint models includes in the linear predictor of the risk model only the random effects of the longitudinal submodel, that is,
\begin{equation}
h_i(t) = h_0(t) \exp ( \bfgamma^\top \bw_i + \bfalpha^\top \bb_i ), \label{Eq:Param-RE}
\end{equation}
where $\bfalpha$ denotes a vector of association parameters each one measuring the association between the corresponding random effect and the hazard for an event. This parameterization is more meaningful when a simple random-intercepts and random-slopes structure is assumed for the longitudinal submodel, in which case the random effects express subject-specific deviations from the average intercept and average slope. Under this setting this parameterization postulates that patients who have a lower/higher level for the longitudinal outcome at baseline (i.e., intercept) or who show a steeper increase/decrease in their longitudinal trajectories (i.e., slope) are more likely to experience the event. In that respect, this formulation shares also similarities with the time-dependent slopes formulation \eqref{Eq:Param-TDslopes}.

A computational advantage of formulation \eqref{Eq:Param-RE} is that it is time-independent, and therefore leads to a closed-form solution (under certain baseline risk functions) for the integral in the definition of the survival function \eqref{Eq:SurvivalFun}. This facilitates computations since we do not have to numerically approximate this integral. However, an important disadvantage of \eqref{Eq:Param-RE} emerges when polynomials or splines are used to capture nonlinear subject-specific evolutions, in which case the random effects do not have a straightforward interpretation, complicating in turn the interpretation of $\bfalpha$. Nonetheless, in our setting, we are primarily interested in predictions and not that much in interpretation, and thus we can consider \eqref{Eq:Param-RE} even under an elaborate mixed model.

%=====================================================

\section{Bayesian Model Averaging} \label{Sec:BMA}
The previous section demonstrated that there are several choices for the link between the longitudinal and event time outcomes. The common practice in prognostic modeling is to base predictions on a single model that has been selected based on an automatic algorithm, such as, backward, forward or stepwise selection, or on likelihood-based information criteria, such as, AIC, BIC, DIC and their variants. However, what is often neglected in this procedure is the issue of model uncertainty. For example, in the scenario that two models are correct, model selection forces us to choose one of the models even if we are not certain which model is true. In addition, with respect to predictions, there could be several competing models that could offer almost equally good predictions or even that some models may produce more accurate predictions for some subjects with specific longitudinal profiles, while other models may produce better predictions for subjects whose profiles have other features. Here we follow another approach and we explicitly take into account model uncertainty by combining predictions under different association structures using Bayesian model averaging (BMA) \citep{hoeting.et.al:99}. We should stress that in our setting there is no concern in using BMA, because we do not average parameters that possibly have different interpretations under the different association structures, but rather predictions that maintain the same interpretation whatever the chosen functional form.

Due to space limitations, we will only focus here on dynamic BMA predictions of survival probabilities. BMA predictions for the longitudinal outcome can be produced with similar methodology. Following the definitions of Section~\ref{Sec:DynPred}, we assume that we have available data $\mathcal D_n = \{T_i, \delta_i, \by_i; i = 1, \ldots, n\}$ based on which we fit $M_1, \ldots, M_K$ joint models with different association structures. Interest is in calculating predictions for a new subject $j$ from the same population who has provided a set of longitudinal measurements $\mathcal Y_j(t) = \{ y_j(s); 0 \leq s \leq t \}$, and has a vector of baseline covariates $\bw_j$. We let $\mathcal D_j(t) = \{\mathcal Y_j(t), T_j^* > t, \bw_j\}$ denote the available data for this subject. The model-averaged probability of subject $j$ surviving time $u > t$, given her survival up to $t$ is given by the expression:
\begin{equation}
\Pr (T_j^* > u \mid \mathcal D_j(t), \mathcal D_n) = \sum \limits_{k = 1}^K \Pr (T_j^* > u \mid M_k, \mathcal D_j(t), \mathcal D_n) \, p(M_k \mid \mathcal D_j(t), \mathcal D_n). \label{Eq:BMA-SurvPred}
\end{equation}
The first term in the right-hand side of \eqref{Eq:BMA-SurvPred} denotes the model-specific survival probabilities, derived as in Section~\ref{Sec:DynPred}, and the second term denotes the posterior weights of each of the competing joint models. The unique characteristic of these weights is that they depend on the observed data of subject $j$, in contrast to classic applications of BMA where the model weights are the same for all subjects. This means that, in our case, the model weights are both subject- and time-dependent, and therefore, for different subjects, and even for the same subject but at different times points, different models may have higher posterior probabilities. Hence, this framework is capable of better tailoring predictions to each subject than standard prognostic models, because at any time point we base risk assessments on the models that are more probable to describe the association between the observed longitudinal trajectory of a subject and the risk for an event.

For the calculation of the model weights we observe that these are written as:
\[
p(M_k \mid \mathcal D_j(t), \mathcal D_n) = \frac{p(\mathcal D_j(t) \mid M_k) \, p(\mathcal D_n \mid M_k) \, p(M_k)}{\sum \limits_{\ell = 1}^K p(\mathcal D_j(t) \mid M_\ell) \, p(\mathcal D_n \mid M_\ell) \, p(M_\ell)},
\]
where
\[
p(\mathcal D_j(t) \mid M_k) = \int p(\mathcal D_j(t) \mid \bftheta_k) p(\bftheta_k \mid M_k) \, d\bftheta_k
\]
and $p(\mathcal D_n \mid M_k)$ is defined analogously. The likelihood part $p(\mathcal D_n \mid \bftheta_k)$ is given by \eqref{Eq:Log-Lik}, whereas $p(\mathcal D_j(t) \mid \bftheta_k)$ equals
\[
p(\mathcal D_j(t) \mid \bftheta_k) = p(\mathcal Y_j(t) \mid \bb_i, \bftheta_k) \, S_j(t \mid \bb_i, \bftheta_k) \, p(\bb_j \mid \bftheta_k).
\]
Thus, the subject-specific information in the model weights at time $t$ comes from the available longitudinal measurements $\mathcal Y_j(t)$ but also from the fact that this subject has survived up to $t$. Closed-form expressions for the marginal densities $p(\mathcal D_n \mid M_k)$ and $p(\mathcal D_j(t) \mid M_k)$ are obtained by means of Laplace approximations \citep{tierney.kadane:86} performed in two-steps, namely, first integrating out the random effects $\bb_i$ and then the parameters $\bftheta_k$. The details are given in the supplementary material. Finally, a priori we assume that all models are equally probable, i.e., $p(M_k) = 1/K$, for all $k = 1, \ldots, K$.

%=====================================================

\section{Analysis of the Aortic Valve Dataset} \label{Sec:Appl}
We return to the Aortic Valve dataset introduced in Section~\ref{Sec:AoValvInf}. Our aim here is to derive a prediction tool that will utilize all recorded information of a patient to provide accurate individualized predictions for both future aortic gradient levels and the risk of re-operation-free survival. Following the discussion of Sections~\ref{Sec:Parms} and \ref{Sec:BMA}, we will compare predictions under different association structures between the longitudinal and survival outcomes, with the BMA predictions that are based on all association structures simultaneously.

We start by defining the set of joint models from which predictions will be calculated. For the longitudinal process we allow a flexible specification of the subject-specific square root aortic gradient trajectories using natural cubic splines of time. More specifically, the linear mixed model takes the form
\begin{eqnarray*}
y_i(t) & = & \beta_1 \mbox{\tt SI}_i + \beta_2 \mbox{\tt RR}_i + \beta_3 \{B_1(t, \lambda) \times \mbox{\tt SI}_i\} + \beta_4 \{B_1(t, \lambda) \times \mbox{\tt RR}_i\}\\
&& \hspace{0.5cm} + \, \beta_5 \{B_2(t, \lambda) \times \mbox{\tt SI}_i\} + \beta_6 \{B_2(t, \lambda) \times \mbox{\tt RR}_i\}\\
&& \hspace{0.5cm} + \, \beta_7 \{B_3(t, \lambda) \times \mbox{\tt SI}_i\} + \beta_8 \{B_3(t, \lambda) \times \mbox{\tt RR}_i\}\\
&& \hspace{0.5cm} + \, b_{i0} + b_{i1}B_1(t, \lambda) + b_{i2}B_2(t, \lambda) + b_{i3}B_3(t, \lambda) + \eps_i(t),
\end{eqnarray*}
where $B_n(t, \lambda)$ denotes the B-spline basis for a natural cubic spline with boundary knots at baseline and 19 years and internal knots at 2.1 and 5.5 years (i.e., the 33.3\% and 66.6\% percentiles of observed follow-up times), {\tt SI}  and {\tt RR} are the dummy variables for the sub-coronary implantation and root replacement groups, respectively, $\eps_i(t) \sim \mathcal N (0, \sigma^2)$ and $\bb_i \sim \mathcal N (\mathbf 0, \bD)$. For the survival process we consider five relative risk models, each positing a different association structure between the two processes, namely:
\begin{eqnarray*}
M_1: && h_i(t) = h_0(t) \exp \bigl \{\gamma_1 \mbox{\tt RR}_i + \gamma_2 \mbox{\tt Age}_i + \gamma_3 \mbox{\tt Female}_i + \alpha_1 m_i(t) \bigl \},\\
M_2: && h_i(t) = h_0(t) \exp \bigl \{\gamma_1 \mbox{\tt RR}_i + \gamma_2 \mbox{\tt Age}_i + \gamma_3 \mbox{\tt Female}_i + \alpha_1 m_i(t) + \alpha_2 m_i'(t)  \bigl \},\\
M_3: && h_i(t) = h_0(t) \exp \bigl \{\gamma_1 \mbox{\tt RR}_i + \gamma_2 \mbox{\tt Age}_i + \gamma_3 \mbox{\tt Female}_i + \alpha_1 \int_0^t m_i(s) ds  \bigl \},\\
M_4: && h_i(t) = h_0(t) \exp \bigl \{\gamma_1 \mbox{\tt RR}_i + \gamma_2 \mbox{\tt Age}_i + \gamma_3 \mbox{\tt Female}_i + \alpha_1 \int_0^t \phi(t - s) m_i(s) ds  \bigl \},\\
M_5: && h_i(t) = h_0(t) \exp \bigl (\gamma_1 \mbox{\tt RR}_i + \gamma_2 \mbox{\tt Age}_i + \gamma_3 \mbox{\tt Female}_i + \alpha_1 b_{i0} + \alpha_2 b_{i1} + \alpha_3 b_{i2} + \alpha_4 b_{i3} \bigl ),
\end{eqnarray*}
where the baseline hazard is approximated with splines, as described in Section~\ref{Sec:JM-submodels}, {\tt Female} denotes the dummy variable for females, and $\phi(\cdot)$ denotes the probability density function of the standard normal distribution. We fitted each of these joint models using a single chain of 115000 MCMC iterations from which we discarded the first 15000 samples as burn-in. All computations have been performed in \R\, (version 2.15.2) and \textsf{JAGS} (version 3.3.0). Trace and auto-correlations plots did not show any alarming indications of convergence failure. Tables~1 and 2 in the supplementary material show estimates and the corresponding 95\% credible intervals for the parameters in the longitudinal and survival submodels, respectively. We observe that the parameter estimates in the relative risk models show much greater variability between the posited association structures than the parameters in the linear mixed models. However, we should note that the interpretation of the regression coefficients $\bfgamma$ is not the same in all five models because we condition on different components of the longitudinal process.

%\citep[][version 2.15.2]{r:12} and \textsf{JAGS} \citep[][version 3.3.0]{plummer:03}

\begin{figure}[!h]
\includegraphics[width = \textwidth]{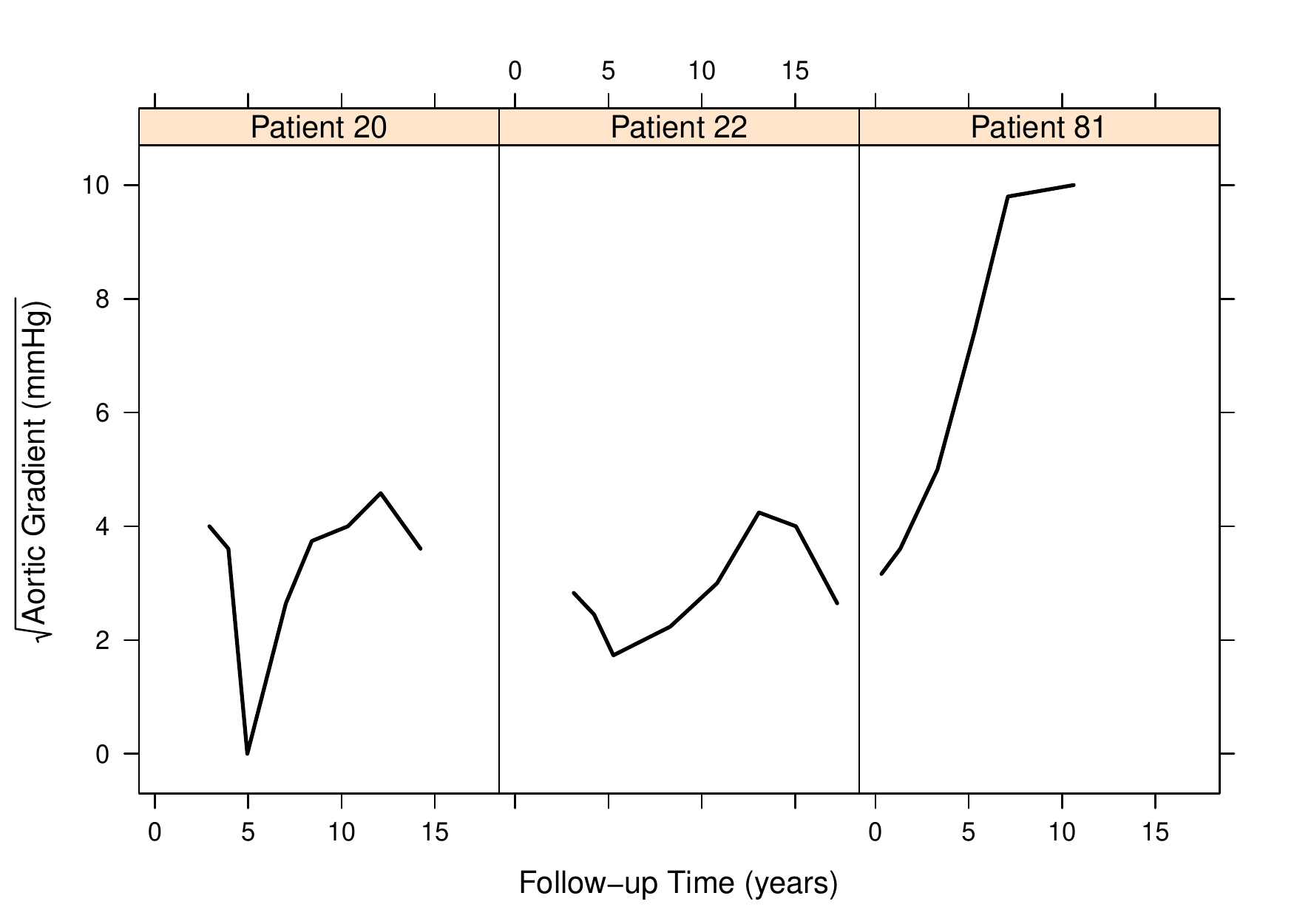}
\caption{Longitudinal trajectories of square root aortic gradient for Patients~20, 22 and 81 who have been excluded from the analysis dataset, and for who we calculate predictions.} \label{Fig:SubSpecProfiles3P}
\end{figure}

\begin{figure}[!h]
\includegraphics[width = \textwidth]{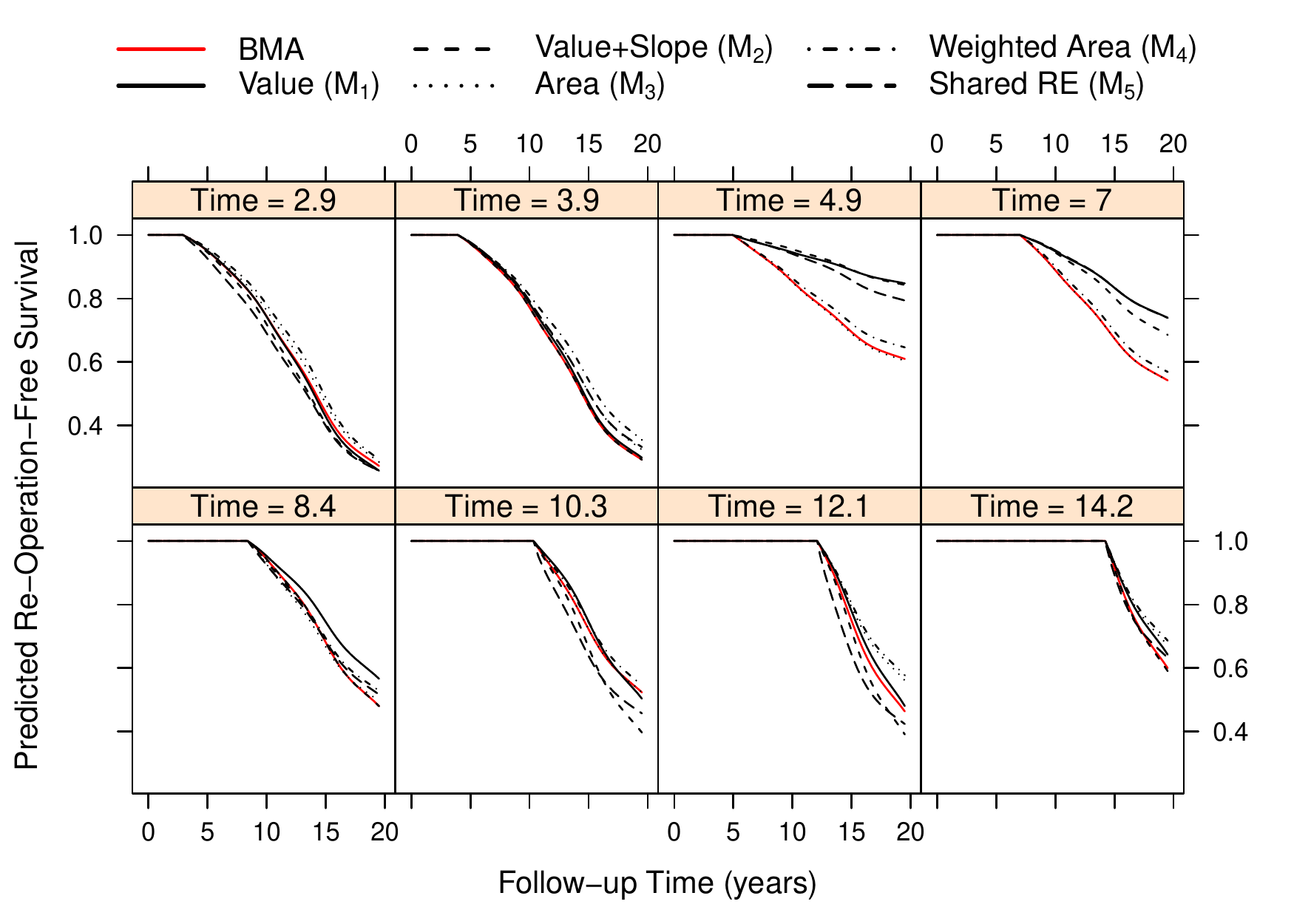}
\caption{Dynamic predictions of re-operation-free survival for Patient 20 under the five joint models along with the BMA predictions. Each panel shows the corresponding conditional survival probabilities calculated after each of her longitudinal measurements have been recorded.} \label{Fig:SurvPredsP20}
\end{figure}

\begin{figure}[!h]
\includegraphics[width = \textwidth]{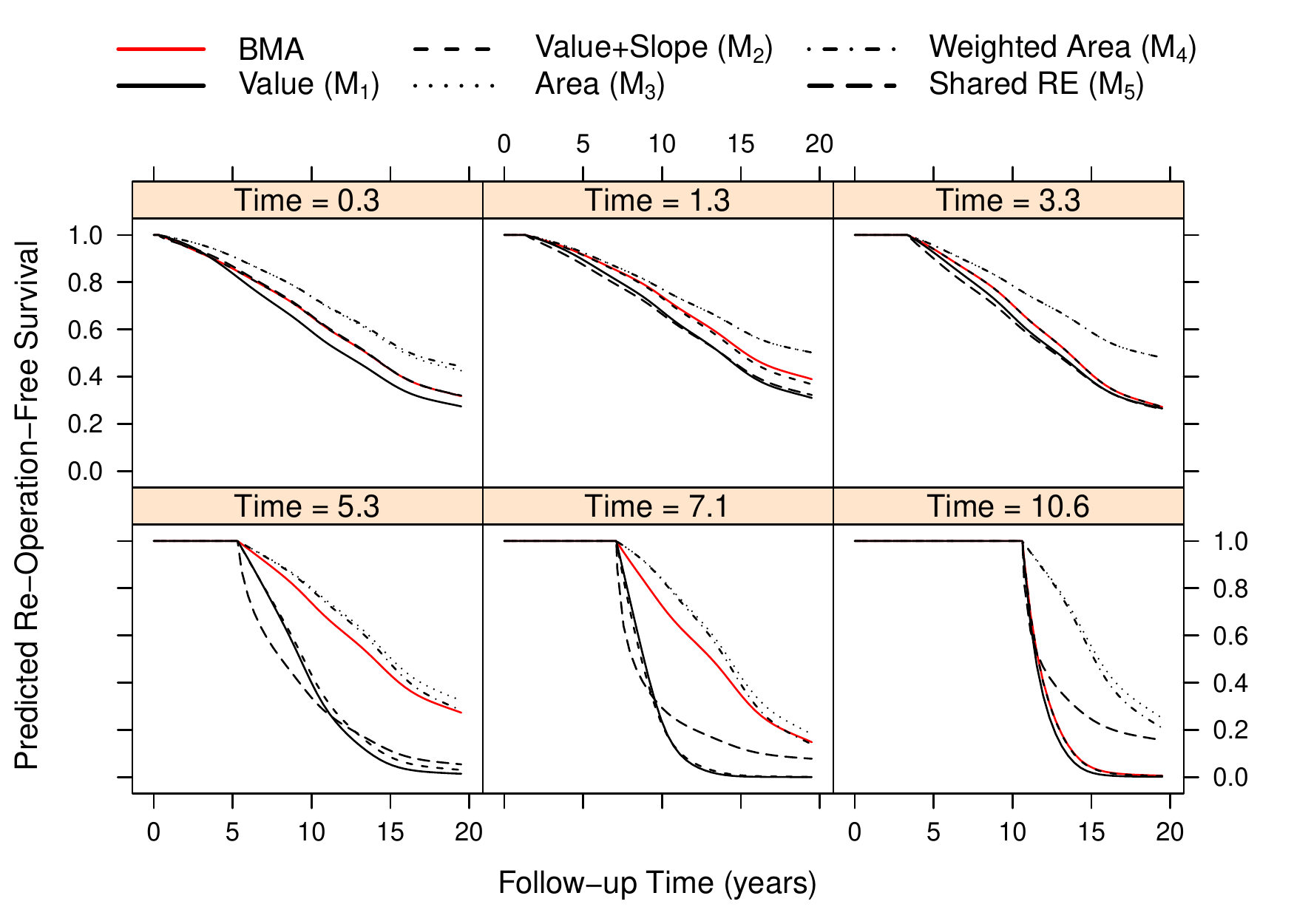}
\caption{Dynamic predictions of re-operation-free survival for Patient 81 under the five joint models along with the BMA predictions. Each panel shows the corresponding conditional survival probabilities calculated after each of his longitudinal measurements have been recorded.} \label{Fig:SurvPredsP81}
\end{figure}

We continue with the calculation of dynamic predictions based on these five joint models. To mimic the real-life use of a prognostic tool, we calculate predictions for three patients, namely Patients 20, 22 and 81, who have been excluded from the dataset we used to fit the joint models. Patient 20 is a female of 64 years old and has received a sub-coronary implantation, Patient 22 is a male of 53 and has also received a sub-coronary implantation, and Patient 81 is a male of 39 and has undertaken a root replacement. The longitudinal trajectories of square root aortic gradient for these patients are illustrated in Figure~\ref{Fig:SubSpecProfiles3P}, from which we see that Patients 20 and 22 show similar profiles with an initial drop in their aortic gradient levels after the operation and up to about five years followed by a stable increase for the next five years and a drop again. On the contrary, Patient 81 showed a steady increase of aortic gradient for the duration of his follow-up. For each one of those subjects we calculate dynamic predictions (i.e., a new prediction after each aortic gradient measurement has been recorded) for both the longitudinal and survival outcomes based on each of the five fitted joint models.
In addition, following the methodology of Section~\ref{Sec:BMA}, we also calculate BMA dynamic predictions for each patient by taking weighted averages of the predictions of the five joint models. We show the predictions of re-operation-free survival probabilities for Patients 20 and 81 in Figures~\ref{Fig:SurvPredsP20} and \ref{Fig:SurvPredsP81}, respectively. The predictions for Patient~22, and the predictions of future square root aortic gradient levels are presented in Figures~1--4 in the supplementary material.
Similarly to the results in Tables~1 and 2 in the supplementary material, we observe that the predictions of survival probabilities seem to be more sensitive to the assumed association structure than the predictions of future square root aortic gradient levels. Much greater differences can be seen for Patient~81 who showed a steeper increase in his aortic gradient levels than the other two patients. Regarding the BMA predictions, it is interesting to note that they show some variability and they are not always dictated by a single joint model. In particular, Table~\ref{Tab:BMAweights} presents the time-dependent subject-specific BMA weights, from which some interesting observations can made.
\begin{table}[ht]
\begin{center}
\begin{tabular}{cccccccc}
\hline
Subject & Year & $(\mbox{Aort.Grad.})^{1/2}$ & $M_1$ & $M_2$ & $M_3$ & $M_4$ & $M_5$ \\
\hline
20 &    2.9 & 4.0 & 0.00 & 0.43 & 0.57 & 0.00 & 0.00 \\
   &    3.9 & 3.6 & 0.00 & 1.00 & 0.00 & 0.00 & 0.00 \\
   &    4.9 & 0.0 & 0.00 & 0.02 & 0.98 & 0.00 & 0.00 \\
   &    7.0 & 2.6 & 0.00 & 0.00 & 1.00 & 0.00 & 0.00 \\
   &    8.4 & 3.7 & 0.00 & 0.99 & 0.01 & 0.00 & 0.00 \\
   &   10.3 & 4.0 & 0.00 & 0.00 & 1.00 & 0.00 & 0.00 \\
   &   12.1 & 4.6 & 0.00 & 0.57 & 0.43 & 0.00 & 0.00 \\
   &   14.2 & 3.6 & 0.00 & 0.88 & 0.12 & 0.00 & 0.00 \\
   \hline
22 &    3.1 & 2.8 & 0.00 & 0.95 & 0.05 & 0.00 & 0.00 \\
   &    4.2 & 2.4 & 0.00 & 1.00 & 0.00 & 0.00 & 0.00 \\
   &    5.3 & 1.7 & 0.00 & 0.00 & 1.00 & 0.00 & 0.00 \\
   &    8.3 & 2.2 & 0.00 & 1.00 & 0.00 & 0.00 & 0.00 \\
   &   10.8 & 3.0 & 0.00 & 1.00 & 0.00 & 0.00 & 0.00 \\
   &   13.1 & 4.2 & 0.00 & 0.01 & 0.99 & 0.00 & 0.00 \\
   &   15.0 & 4.0 & 0.00 & 1.00 & 0.00 & 0.00 & 0.00 \\
   &   17.3 & 2.6 & 0.00 & 0.67 & 0.33 & 0.00 & 0.00 \\
   \hline
81 &    0.3 & 3.2 & 0.00 & 1.00 & 0.00 & 0.00 & 0.00 \\
   &    1.3 & 3.6 & 0.00 & 0.84 & 0.16 & 0.00 & 0.00 \\
   &    3.3 & 5.0 & 0.00 & 1.00 & 0.00 & 0.00 & 0.00 \\
   &    5.3 & 7.4 & 0.00 & 0.17 & 0.83 & 0.00 & 0.00 \\
   &    7.1 & 9.8 & 0.00 & 0.19 & 0.81 & 0.00 & 0.00 \\
   &   10.6 & 10.0 & 0.00 & 0.99 & 0.01 & 0.00 & 0.00 \\
   \hline
  \end{tabular}
\caption{BMA posterior weights for the five joint models for each subject and after each measurement.}
\label{Tab:BMAweights}
\end{center}
\end{table}
First, only models $M_2$ and $M_3$ contribute in the BMA predictions of these three patients with the weights of the other three models being practically zero. Second, we indeed observe that the choice of the most appropriate model changes in time; for example, for Patient~20 and using only the first measurement we observe that there is little to choose between models $M_2$ and $M_3$, when the second measurement is observed then $M_2$ dominates, whereas when the third measurement is recorded prediction is based solely on model $M_3$ (though predictions based on the first two measurements are very similar from all five models, more variability is observed for latter time points). Similar behavior is observed for the other two patients as well. This convincingly demonstrates that it would be not optimal to base predictions on a single model for all patients and during the whole follow-up.

%=====================================================

\section{Simulations} \label{Sec:Simul}
\subsection{Design} \label{Sec:Simul-Design}
We performed a series of simulations to evaluate the finite sample performance of the BMA subject-specific predictions. The design of our simulation study is motivated by the set of joint models fitted to the Aortic Valve dataset in Section~\ref{Sec:Appl}. In particular, we assume 300 patients who have been followed-up for a period of 19 years, and were planned to provide longitudinal measurements at baseline and afterwards at nine random follow-up times. For the longitudinal process, and similarly to the model fitted in the Aortic Valve dataset, we used natural cubic splines of time with two internals knots placed at 2.1 and 5.5 years, and boundary knots placed at baseline and 19 years, i.e., the form of the model is as follows
\begin{eqnarray*}
y_i(t) & = & \beta_1 \mbox{\tt Trt0}_i + \beta_2 \mbox{\tt Trt1}_i + \beta_3 \{B_1(t, \lambda) \times \mbox{\tt Trt0}_i\} + \beta_4 \{B_1(t, \lambda) \times \mbox{\tt Trt1}_i\}\\
&& \hspace{0.5cm} + \, \beta_5 \{B_2(t, \lambda) \times \mbox{\tt Trt0}_i\} + \beta_6 \{B_2(t, \lambda) \times \mbox{\tt Trt1}_i\}\\
&& \hspace{0.5cm} + \, \beta_7 \{B_3(t, \lambda) \times \mbox{\tt Trt0}_i\} + \beta_8 \{B_3(t, \lambda) \times \mbox{\tt Trt1}_i\}\\
&& \hspace{0.5cm} + \, b_{i0} + b_{i1}B_1(t, \lambda) + b_{i2}B_2(t, \lambda) + b_{i3}B_3(t, \lambda) + \eps_i(t),
\end{eqnarray*}
where $B_n(t, \lambda)$ denotes the B-spline basis for a natural cubic spline, {\tt Trt0} and  {\tt Trt1} are the dummy variables for the two treatment groups, $\eps_i(t) \sim \mathcal N (0, \sigma^2)$ and $\bb_i \sim \mathcal N (\mathbf 0, \bD)$ with $\bD$ taken to be diagonal.

For the survival process, we have assumed four scenarios, each one corresponding to a different functional form for the association structure between the two processes. More specifically, \begin{eqnarray*}
\mbox{Scenario I:} && h_i(t) = h_0(t) \exp \bigl \{\gamma_0 + \gamma_1 \mbox{\tt Trt1}_i + \alpha_1 m_i(t) \bigl \},\\
\mbox{Scenario II:} && h_i(t) = h_0(t) \exp \bigl \{\gamma_0 + \gamma_1 \mbox{\tt Trt1}_i + \alpha_1 m_i(t) + \alpha_2 m_i'(t)  \bigl \},\\
\mbox{Scenario III:} && h_i(t) = h_0(t) \exp \bigl \{\gamma_0 + \gamma_1 \mbox{\tt Trt1}_i + \alpha_1 \int_0^t m_i(s) ds  \bigl \},\\
\mbox{Scenario IV:} && h_i(t) = h_0(t) \exp \bigl (\gamma_0 + \gamma_1 \mbox{\tt Trt1}_i + \alpha_1 b_{i0} + \alpha_2 b_{i1} + \alpha_3 b_{i2} + \alpha_4 b_{i3} \bigl ),
\end{eqnarray*}
with $h_0(t) = \sigma_t t^{\sigma_t - 1}$, i.e., the Weibull baseline hazard. The values for the regression coefficients in the longitudinal and survival submodels, the variance of the error terms of the mixed model, the covariance matrix for the random effects, and the scale of the Weibull baseline risk function are given in Section~3 of the supplementary material. Censoring times were simulated from a uniform distribution in $(0, t_{max})$ with $t_{max}$ set to result in about 45\% censoring in each scenario. For each scenario we simulated 200 datasets.

%%%%%%%%%%%

\subsection{Results} \label{Sec:Simul-Results}
Motivated by the Aortic Valve dataset to produce accurate risk predictions, but also from previous experience regarding the robustness of predictions for the longitudinal outcome on the assumed association structure, we have focused on dynamic predictions for the survival outcome. More specifically, under each scenario and for each simulated dataset, we randomly excluded ten subjects whose event times were censored (it is more meaningful to calculate predictions for those individuals since they have not experienced the event yet), and in the remaining patients, we fitted five joint models. The longitudinal submodel was the same as the one we simulated from. For the survival process we assumed relative risk submodels with: the current value term (as in Scenario I), the current value and current slope terms (as in Scenario II), the cumulative effect (as in Scenario III), the weighted cumulative effect \eqref{Eq:Param-wCumEff} with weight function the probability density function of the standard normal distribution, and only including the random effects (as in Scenario IV). Because the aim of this simulation study was to investigate how the different association structures affect predictions, we also assumed the same Weibull baseline hazard function from which we simulated the data and not the B-spline approximated baseline hazard \eqref{Eq:BaseHaz} presented in Section~\ref{Sec:JM-submodels}.

\begin{figure}[!h]
\includegraphics[width = \textwidth]{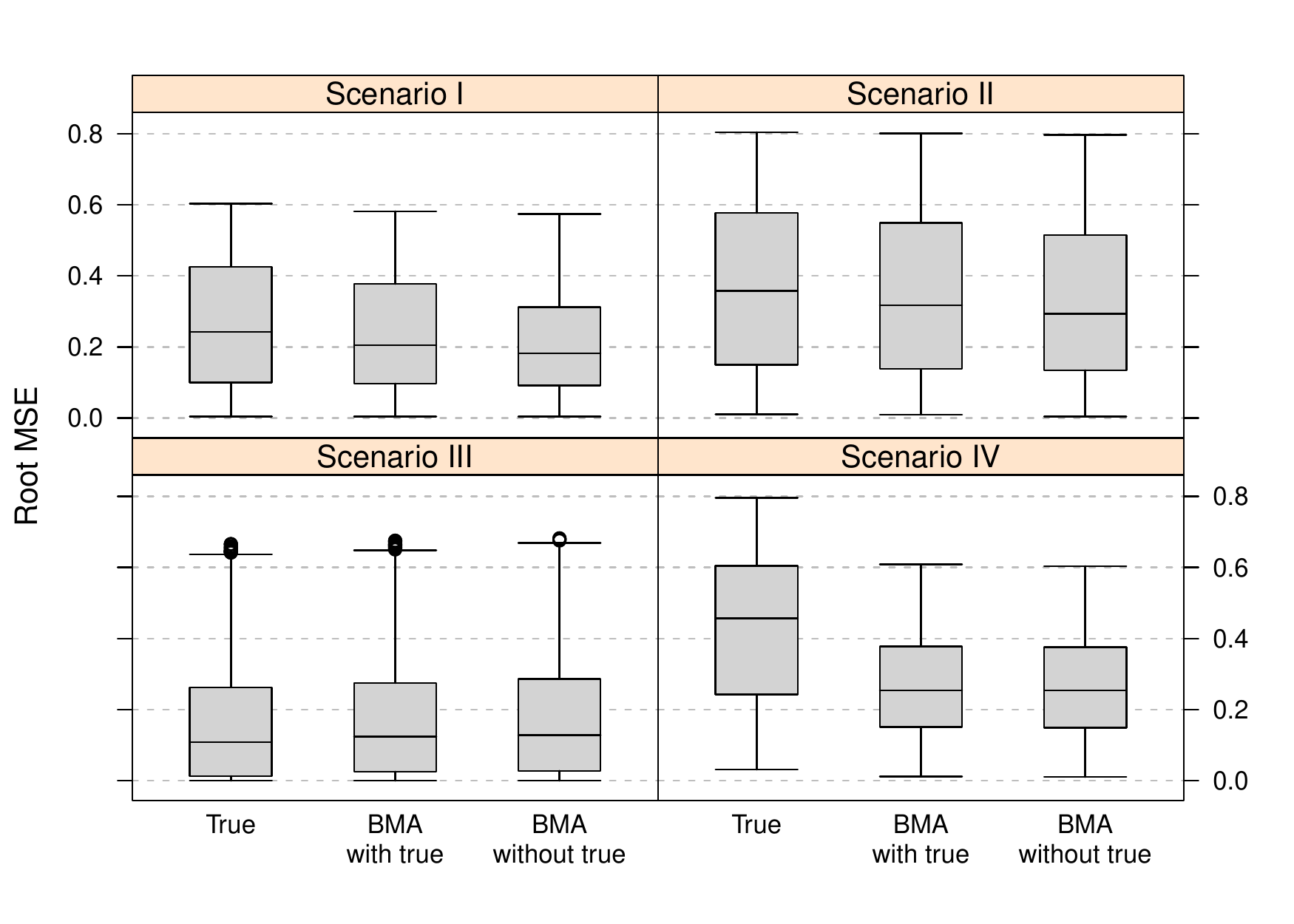}
\caption{Simulation results under the four scenarios based on 200 datasets. Each boxplot shows the distribution of the root mean squared predictions error of the corresponding method to compute predictions (i.e., based on the true model, BMA including the true model and BMA without the true model) versus the gold standard.} \label{Fig:SimulResults}
\end{figure}

Based on the five fitted joint models, we calculated predictions for each of the ten subjects we have originally excluded, at ten equidistant time points between their last available longitudinal measurement and the end of follow-up. These predictions were calculated under the true model for each scenario, BMA including all five models, and BMA based on four models without including the true one for the specific scenario. These predicted survival probabilities were then compared with the gold standard survival probabilities, calculated as $S_j \bigl \{ u \mid \mathcal M_j(u, \bb_j), \bftheta \bigr \} / S_j \bigl \{ t \mid \mathcal M_j(t, \bb_j), \bftheta \bigr \}$, using the true parameter values and the true values of the random effects for the subjects we excluded. In each simulated dataset and for each of the ten subjects, we calculated root mean squared prediction errors (RMSEs) between the gold standard survival probabilities and the predictions under the three methods. The RMSEs over all the subjects from the 200 datasets are shown in Figure~\ref{Fig:SimulResults}.
For all scenarios we observe that in practice the BMA subject-specific predictions perform very well against the corresponding predictions from the true model. The greatest differences are observed in Scenario IV in which the BMA predictions seem to considerably outperform the true model predictions. A more careful examination of this scenario showed that this behavior was due to the fact that all models produced on average accurate predictions, but the predictions from the true model were much more variable than ones from the other models. In addition, for the scenarios we considered in this study, it seems that BMA works equally well whether or not the true model is included in the list of the models that are averaged. This is a promising result because it suggests that BMA-based predictions could `protect' against a misspecification of the association structure between the two processes.

%=====================================================

\section{Discussion} \label{Sec:Disc}
This paper illustrated how dynamic predictions from joint models with different association structures can be optimally combined using Bayesian model averaging. The novel feature of this approach is that the weights for combing predictions depend on the recorded information for the subject for whom predictions are of interest. Thus, for different subjects and even for the same subject but at different follow-up times, different models may have higher weights. This explicitly accounts for model uncertainty and acknowledges that a single prognostic model may not be adequate for quantifying the risk of all patients. Our simulation study showed that BMA predictions perform very well in comparison with predictions from the true model, even if the true model is not included in the list of models that are averaged. This gives us more confidence in trusting BMA for deriving predictions for future patients from the Aortic Valve study population based on the five joint models we have considered.

In our developments we have considered a simple joint model for a single longitudinal outcome and one time-to-event. However, often in longitudinal studies several outcomes are recorded on each patient during follow-up. For example, for the patients enrolled in the Aortic Valve study also aortic regurgitation was recorded, which is another measure of valve function. Hence it would be of interest to investigate whether by considering both longitudinal biomarkers we could improve the accuracy of predictions. In addition, in our analysis we have considered the composite event re-operation or death (whatever comes first), but for the treating physicians it could be of interest to have separate risk estimates for the two events. Based on recent advances in joint modeling that include joint models for multiple markers and multiple event times \citep{huang.et.al:11, rizopoulos.ghosh:11, liu.huang:09}, our ideas could be relatively easily extended to these more elaborate cases. The challenge in such settings will be the high dimensionality of the model space. This is because the number of possible combinations of association structures between the longitudinal and the survival processes grows exponentially with the number of outcomes.

A topic that we have partially addressed in this paper, but which is of high relevance for the practicality of prognostic models concerns the external validation of the derived predictions in terms of both discrimination and calibration. In particular, from our simulations we saw that the BMA predictions perform satisfactorily compared to the predictions using the true model, but this does not answer the question of how accurately the longitudinal outcome can predict the survival one or if it can discriminate between subjects of low and high risk. In standard survival analysis there has been a lot of research at these two fronts \citep[see e.g.,][and references therein]{zheng.heagerty:07, gerds.schumacher:06}, but within the joint modeling framework relatively little work has been done \citep{rizopoulos:11, proust-lima.taylor:09}. Theoretically, all standard measures for calibration (e.g., Brier score) and discrimination (time-dependent sensitivity, specificity and ROC curves) can be defined for joint models, but their estimation is more challenging.

%Finally, to facilitate the practicality of our proposals, we have written \R\, and %\textsf{JAGS} code that fits joint models for continuous longitudinal outcomes and event %times under the association structures described in Section~\ref{Sec:Parms}, and also %implements the BMA predictions of Section~\ref{Sec:BMA}. More information is available upon %request by the authors.

%=====================================================

\section{Supplementary Material}
Supplementary material are available in \texttt{SuppPredParam.pdf}, and include \textbf{Section 1}: Figures and Tables with results from the analysis of the Aortic Valve dataset; \textbf{Section 2}: Details on the Laplace approximations to calculate $p(\mathcal D_n \mid M_k)$ and $p(\mathcal D_j(t) \mid M_k)$; \textbf{Section 3}: Details on the simulation settings.

%=====================================================

\bibliographystyle{asa}
\bibliography{predParam}

\end{document}